# Preliminary results of Resistive Plate Chambers operated with eco-friendly gas mixtures for application in the CMS experiment

______________________________________________________________________________

M. Abbrescia[a], P. Van Auwegem[b], L. Benussi[c], S. Bianco[c], S. Cauwenbergh[b], M. Ferrini[d], S. Muhammad[cd], L. Passamonti[c], D. Pierluigi[c], D. Piccolo[c,*], F. Primavera[c], A. Russo[c], G. Saviano[cd], M. Tytgat[b]

[a] *INFN, Sezione di Bari, Via Prabona 4, IT-70126 Bari, Italy*
[b] *Ghent University, Dept. of Physics and Astronomy, Proeftuinstraat 86, B-9000 Gent, Belgium*
[c] *Laboratori Nazionali di Frascati dell'INFN, Via E. Fermi, 40, 00044 Frascati (Roma) Italy*
[d] *Laboratori Nazionali di Frascati dell'INFN and Facolta' di Ingegneria Roma1, Via Eudossiana 18 00184 (Roma) Italy.*

E-mail: dpiccolo@lnf.infn.it

ABSTRACT: The operations of Resistive Plate Chambers in LHC experiments require Fluorine based (F-based) gases for optimal performance. Recent European regulations demand the use of environmentally unfriendly F-based gases to be limited or banned. In view of the CMS experiment upgrade, several tests are ongoing to measure the performance of the detector with these new ecological gas mixtures, in terms of efficiency, streamer probability, induced charge and time resolution. Prototype chambers with readout pads and with the standard CMS electronic setup are under test. In this paper preliminary results on performance of RPCs operated with a potential eco-friendly gas candidate 1,3,3,3-Tetrafluoropropene, commercially known as HFO-1234ze, with $CO_2$ and $CF_3I$ based gas mixtures are presented and discussed for the possible application in the CMS experiment.

---

[*] Corresponding author.

# Contents

# 1. Introduction

Resistive Plate Chambers (RPCs)[1] are widely used in High Energy Physics applications and in particular in the LHC experiments.

For applications where high background rates are expected, they have to be operated in avalanche mode in order to keep the total produced charge low with benefits in terms of aging and rate capability. This is usually obtained with suitable gas mixtures that prevents the transition from avalanche to streamer modes keeping the detection efficiency above 90%. The use of Fluorine (F)-based gases, usually used in refrigerants, have shown so far to give the best performance.

The RPC system of the Compact Muon Solenoid (CMS)[2] experiment is operated with a gas mixture of $C_2H_2F_4$ 95.2%, isobuthane 4.5% and $SF_6$ 0.3% showing very high performance in the LHC environment[3].

Recent European regulations demand the use of environmentally un-friendly F-based gases to be limited or banned. The impact of a refrigerant on the environment has to be quantified in terms of the contribution to the greenhouse effect and to the depletion of the ozone layer. The first mentioned effect is measured in Global Warmth Potential (GWP), and is normalized to the effect of $CO_2$ (GWP = 1), and the effect on the ozone layer is measured in Ozone Depletion Potential (ODP), normalized to the effect of $CCl_3F$ (ODP = 1). The European Community has prohibited the production and use of gas mixtures with GWP > 150.

The $C_2H_2F_4$ and $SF_6$ gases used in the RPCs for example present a GWP=1430 and 23900 respectively and need to be replaced with components with lower GWP. In the last two years a program of measurements has been started inside the CMS Collaboration in order to find the right eco-friendly candidate given the constrains defined by the front end electronic used in the CMS experiment. An overview of potential gas candidates can be found in[4].

In this paper we will concentrate on the use of HFO1234ze and $CF_3I$ component in the RPC gas mixture. First results of RPCs operated with this component and in streamer mode have been already presented in the past[5] and results in avalanche mode were presented by this group in[6].
This paper reports the new tests performed and additional gas combination investigated.

## 2. Experimental Setup

Studies of new ecological gas mixtures for the CMS experiment are carried out in two different laboratories: INFN National Laboratories of Frascati and at Ghent University.

Two complementary approaches are followed in the two laboratories.

In Frascati an array composed of 12 single-gap square (50 x 50) cm$^2$ RPC detectors is in operation and arranged as a cosmic ray hodoscope. The system is a copy of the Gas Gain Monitoring (GGM) system of the CMS RPC muon detector at the Large Hadron Collider (LHC) of CERN, Geneva, Switzerland described elsewhere [7][8].

The system is located inside a temperature and relative humidity controlled box and flushed with different gas mixtures humidified at about 40% (fig. 1).

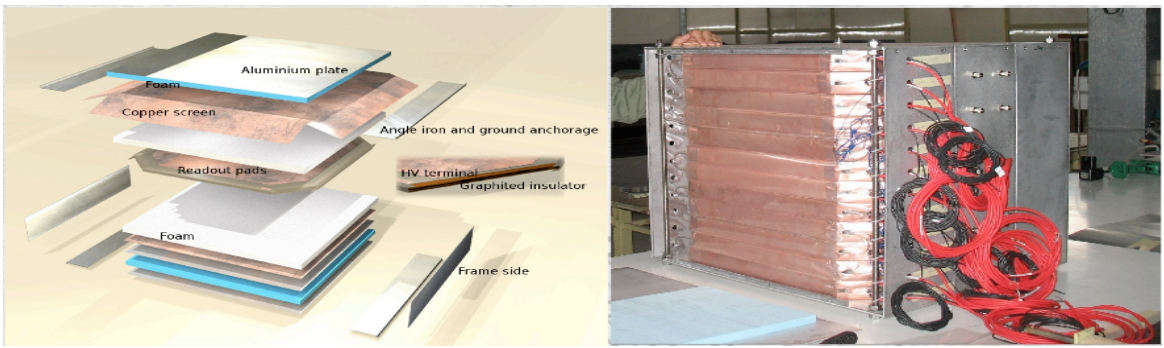

**Figure 1: Left: layout of the single gap RPC with double pad readout used for tests; Right: setup of the RPC tower.**

Each chamber of the GGM system consists of a single gap with double-sided pad read-out: two copper pads are glued on the two opposite external sides of the gap. Two foam planes are used to reduce the capacity coupling between the pad and the copper shields used as screen for the system. The signal is read-out by a transformer based circuit that allows to algebraically subtract the two signals, which have opposite polarities, and to obtain an output signal with subtraction of the coherent noise, with an improvement of about a factor 4 of the signal to noise ratio.

The trigger is provided by scintillator layers located on the top and bottom of the RPC stack with an average event rate of about 1 Hz, corresponding to about 20 minutes for 1000 events. The induced signal on the pad is acquired by a 5 GSample bandwidth oscilloscope and used to measure efficiency, streamer probability and time resolution as a function of Effective Voltage.

More details on the experimental setup and on the analysis criteria can be found in [6] and [9].

In Ghent, two standard CMS-like double-gap chambers are operational and tested with cosmic rays using the same front-end electronics used in the CMS experiment.

The advantage of this approach is to measure the efficiency and cluster size for different gas mixtures with a setup identical to the one used inside the CMS experiment. On the contrary the first approach allows collecting charge spectra and time distributions giving the possibility to extract with a dedicated analysis the efficiency and the streamer probability.

In this paper will be presented mainly results coming from the first approach of Frascati group plus some preliminary result from the Ghent setup.

The plans for the future are to improve the synergy between the two approaches.

## 3. Experimental Results

In this section an overview of the main results will be reported. Many different gas mixtures have been tested and for each of them the efficiency, streamer probability, induced charge and time resolution as a function of the effective Voltage (normalized to $20^o$c and 990 mbar) are reported.

The detector is considered efficient if the induced charge is above 300 fC and the peak voltage is above 0.4 mV. These thresholds are very similar to the threshold used with the standard CMS front-end electronics. A signal is considered a streamer if the induced integrated charge exceed 40 pC.

For the results taken with the Ghent station, the standard CMS electronics is used and there is no ambiguity in the efficiency definition.

### 3.1 Replacing tetrafluoroethane with tetrafluoropropane

The quenching power of the tetrafluoropropane is much higher than the tetrafluorethane. Replacing all the R134a with HFO-1234ze cause the working point to be moved to values much higher than the power system capabilities. In order to check the performance we replaced just partially the amount of R134a in the gas mixture.

Figs. 2 and 3 show the efficiency and streamer probability, induced charge and time resolution as a function of the effective Voltage, respectively, for the new gas mixtures compared to the standard CMS gas mixture. As can be seen replacing 45% of R134a with HFO1234-ze move already the working point to values above 13 KVolts.

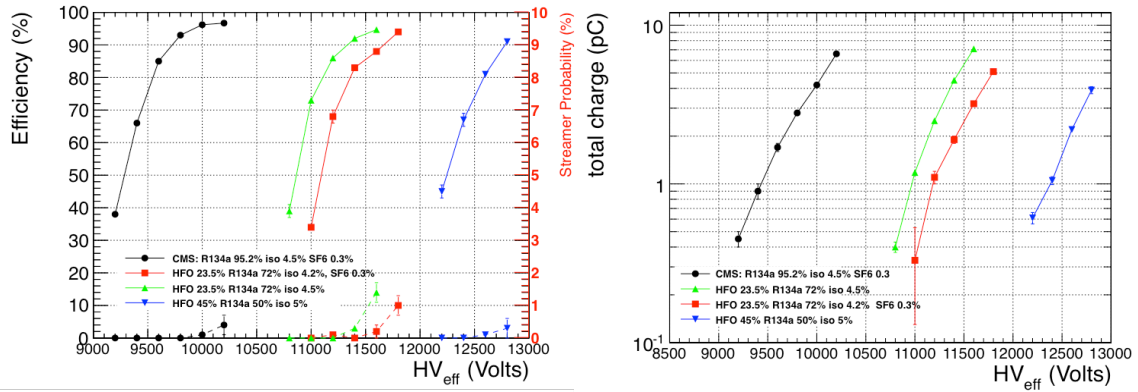

**Figure 2:** Left: Efficiency (solid line) and streamer probability (dashed line) as a function of effective voltage for HFO-1234ze - R134a based gas mixtures; Right: Integrated charge vs. effective voltage for HFO1234ze-r134a based gas mixtures (due to the double pad readout, the values should be divided by ~2 if compared with single pad or strip readout).

To keep the working voltage at lower values a first attempt has been done replacing tetrafluorethane with Argon and keeping the tetrafluoropropane as main component. Fig. 4 show the results in terms of efficiency and streamer probability vs effective voltage. Too low fractions of Argon are not enough to work at values below 10 kV, while increasing the percentage of Argon suddenly trigger streamers at rates not acceptable for LHC experiments.

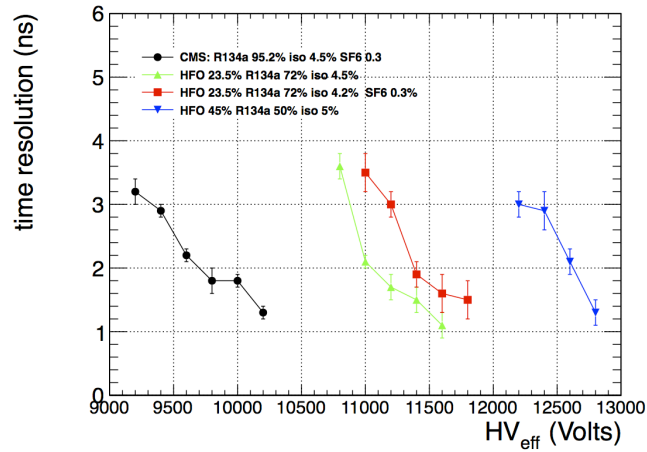

**Figure 3:** Time resolution as a function of effective voltage for HFO1234ze-r134a based gas mixtures.

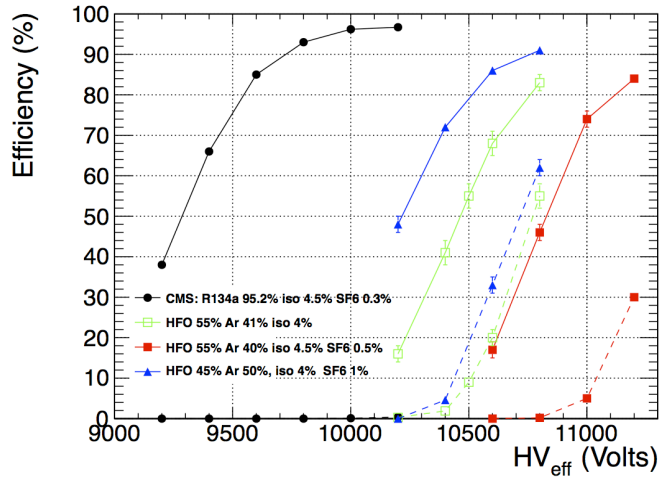

**Figure 4:** Efficiency (solid line) and streamer probability (dashed line) for HFO1234ze-Argon based gas mixtures.

### 3.2 CO$_2$-tetrafluoropropane based gas mixtures

Mixtures based on CO$_2$ and tetrafluoropropane have been tested in order to limit the working voltage and keep under control the streamer probability.

Figures 5 and 6 left, show the results for different fractions of CO$_2$/HFO-1234ze keeping constant around 4% for the iso-buthane and around 1% for the SF$_6$. Results are quite interesting in particular for a fraction of 31% of tetrafluoropropane (green upper pointing triangles in the figure). Although the working point is above 10 kV, the streamer probability seems to be under control. The same results have been cross-checked with the Ghent test station in figure 6 (right), in which the working point is a bit lower, but we have to consider the double gap design of the Ghent chambers and a small difference in the gas mixture.

The CO$_2$ and tetrafluoropropane based gas mixtures seem a good starting point towards a good ecological gas mixtures working in avalanche mode.

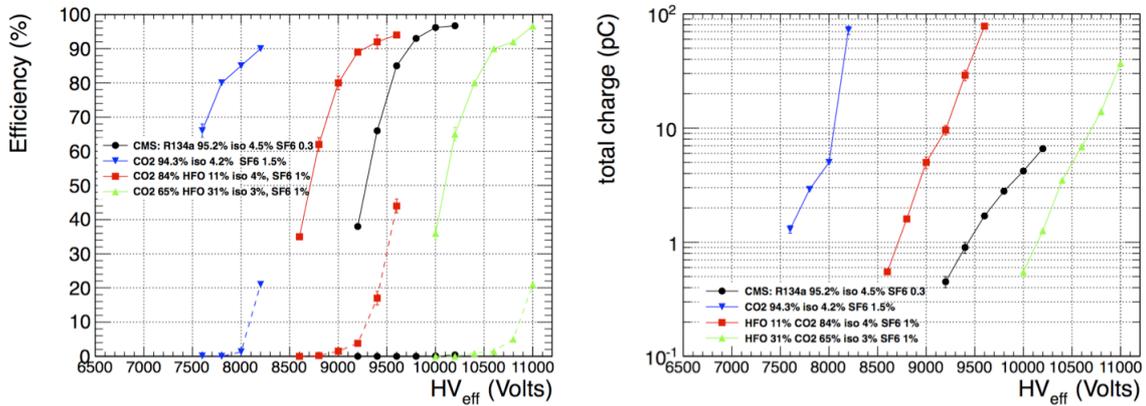

**Figure 5**: Left: Efficiency (solid line) and streamer probability (dashed line) as function of effective voltage, for $CO_2$ based gas mixtures; Right: Integrated charge vs. effective voltage for $CO_2$ based gas mixtures (due to the double pad readout, the values should be divided by ~2 if compared with single pad or strip readout).

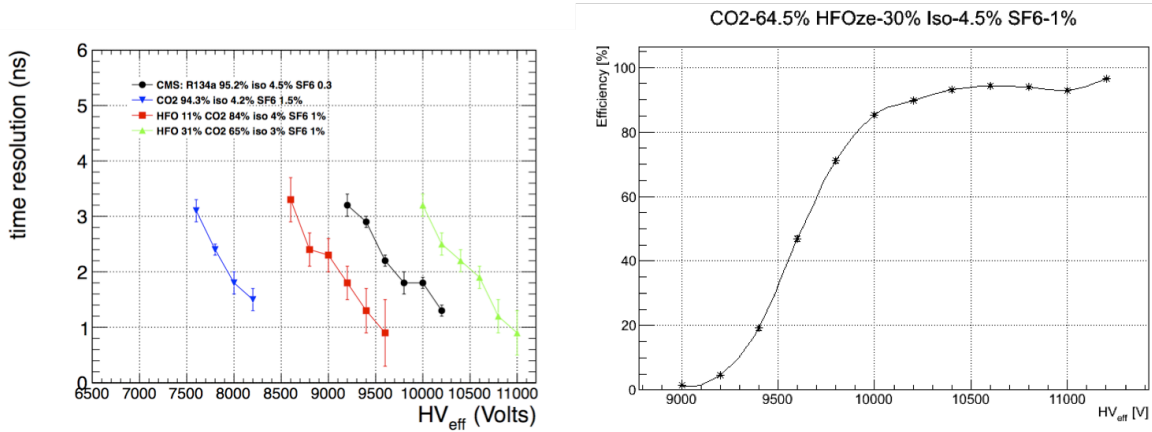

**Figure 6**: Left: Time resolution as function of effective voltage for $CO_2$ based gas mixtures; Right: Efficiency vs. Effective Voltage for a CO2 based gas mixture tested at Ghent Laboratory.

### 3.3 Helium-tetrafluoropropane based gas mixtures

Helium is a gas characterized by no vibrational and rotational degrees of freedom and with a large first ionization energy, being around 24.6 eV, much higher than tetrafluoroethane or tetrafluoropropane. As a consequence the Helium in first approximation does not take part in the avalanche process and just act as a "place holder" reducing the partial pressure of the gas mixture and so moving the working voltage to lower values. So, in presence of gas mixtures characterized by a high operating voltage, it is a natural idea to add a fraction of Helium to lower the operating voltage while preserving stable performance.

This has already been proved some years ago [10] with quite encouraging results.

Now we tested the use of the Helium in addition to tetrafluoropropane based gas mixtures. Complete set of results and details of the measurements can be found in [9].

In fig. 7 are shown the most interesting results in terms of efficiency and streamer probability for purely ecological gas mixtures working at reasonable operational voltage.

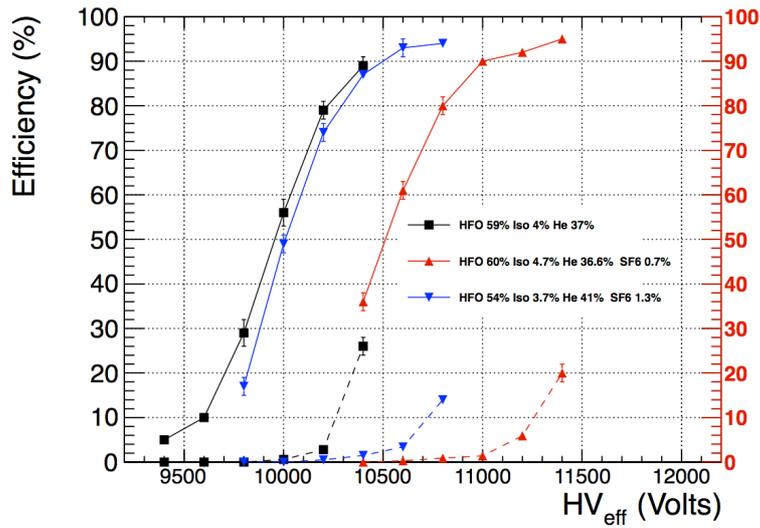

**Figure 7**: Efficiency (solid line) and streamer probability (dashed line) as function of effective voltage, for Helium based gas mixtures.

### 3.4 $CF_3I$ based gas mixtures

A different component that present chemical characteristics suitable for RPC operations is the Trifluoroiodomethane ($CF_3I$). This gas has a Global Warming Power and an Ozone Depletion Power close to zero. The disadvantage of this gas is at moment the price very much higher than tetrafluorethane and tetrafluoropropane.

A small quantity of this gas has been bought at Frascati laboratory and used for very preliminary tests. It shows a much higher quenching power with respect to HFO-1234ze so that only very small fractions can be added in the gas mixtures. The figures 8 (left) and (right) show the efficiency and streamer probability for tetrafluorethane-$CF_3I$ and $CO_2$-$CF_3I$ based gas mixtures respectively.

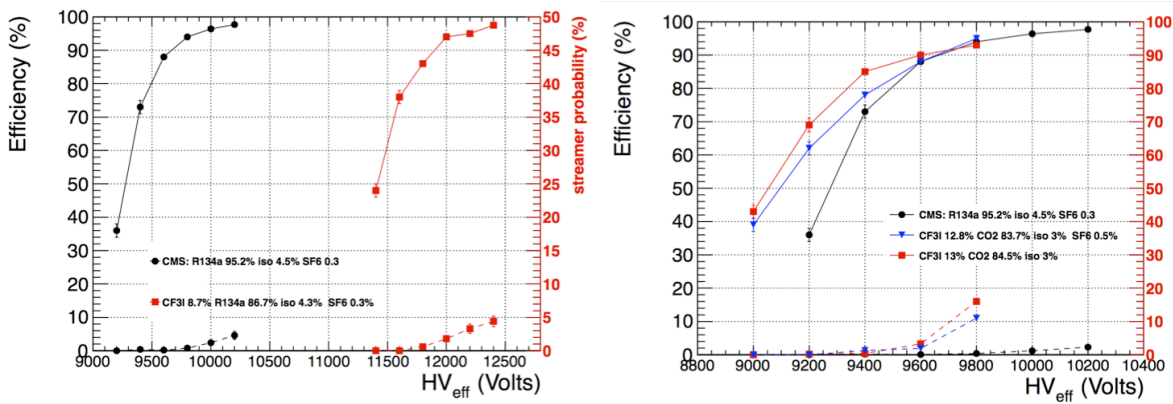

**Figure 8**: Efficiency (solid line) and streamer probability (dashed line) for $CF_3I$-r134a based gas mixtures (left) and $CF_3I$-$CO_2$ based gas mixtures (right).

## 4. Conclusions

The problem to replace the tetrafluoroethane with new ecological components for the operations of the RPC in the CMS experiment is not trivial. Due to the constraint defined by the present front end electronics and High Voltage system, it is not easy to find an ecological mixtures reproducing the same performance as the standard CMS gas mixture.

Many eco-friendly gas mixtures have been tested in both Frascati and Ghent laboratories. The full replacement of tetrafluoroethane with tetrafluoropropane is not possible due to very high working voltage needed for the RPC operations. Many alternatives have been tried adding Argon, $CO_2$, Helium in different percentages, or other gases like $CF_3I$. Results are encouraging but still more work is needed to find the best solution for the CMS purposes.